\documentclass[10pt]{article}%

\usepackage{amssymb}
\usepackage{latexsym}
\usepackage{amsmath}
\usepackage{amsthm}
\usepackage{graphicx}
\usepackage{fancyhdr}

\def\P{\mathbb{P}}

\def\ceil#1{\lceil{#1}\rceil} 

\theoremstyle{plain}

\theoremstyle{definition}

\def\cover{
\centerline{\large Effects of multiple gene control on the spread of altruism by
group selection}
\vskip.5cm

\centerline{Tom\' a\v s Kulich$^*$, Jaroslav Flegr$^\dag$}
\vskip1cm
$^*$Department of Computer Science, Faculty of Mathematics, Physics and
Informatics, Comenius University, Mlynsk\' a dolina, 84248 Bratislava, Slovak
Republic. E-mail: kulich@dcs.fmph.uniba.sk
\vskip0.5cm
$^\dag$Department of Philosophy and History of \linebreak[0] Science, Charles
University, \linebreak Vini\v cn\' a~7, CZ-128 44 Praha 2, Czech Republic. E-mail: flegr@cesnet.cz
\vskip1cm
\centerline{\textbf{Abstract}}
The origin of altruistic behavior, i.e. the behavior that is useful for a
population or a species but goes at the expense of an altruistic individual, has
long been a challenge for students of evolutionary biology. The populations with
altruistic individuals thrive better than those without altruists; however, the
altruists within a population thrive worse than the non-altruists and their
prevalence in the population decreases due to individual selection. Under
certain conditions, the strength of group selection, i.e. the competition
between populations, can surpass the strength of individual selection; however,
such conditions are rarely achieved in practice. It was suggested recently that
chances for altruistic behavior to spread highly increase when it is controlled
not by a single gene but by multiple independent genes substitutable in their
effects on the phenotype of the individual. Here we confirm the original verbal
model published as part of the frozen plasticity theory by numerical modeling of
the spread of altruistic/selfish alleles in a metapopulation consisting of
partly isolated groups of organisms (demes) interconnected by migration. We have
shown that altruistic behavior coded by multiple substitutable genes can stably
coexist with selfish behavior, even under relatively high mutation and migration
rates, i.e. under such conditions where altruistic behavior coded by a single
gene is quickly outcompeted in a metapopulation.  }

\begin{document}
\cover
 
\section{Introduction}  
The problem of evolution and persistence of altruistic behavior has long been a
challenge for theoretical and evolutionary biology. According to classical
models, a behavioral pattern that provides an advantage to a group and at the
same time places its carrier at a disadvantage has a low chance of spreading and
enduring in nature. Groups in which the altruistic trait spreads would thrive
better than those in which this trait is lacking and the average fitness of
their members would be greater; however, selfish individuals who do not exhibit
this trait and do not behave altruistically, but only enjoy the advantages
provided by the presence of altruists, would have the greatest fitness within
these groups. It has been shown that under certain conditions, the strength of
group selection can surpass that of individual selection, especially in
populations with a certain structure and certain population dynamics \cite{4}.
However, most analyses have shown that under usual conditions, the spreading of
an altruistic behavioral pattern is rather rare.

The chances for altruistic behavior to spread may considerably increase when
complex gene interactions are responsible for the altruistic behavior. For
example, the individuals behave altruistically when heterozygous in a particular
altruistic gene while behaving selfishly when homozygous in such a gene or
alternatively altruistic behavior is coded by multiple independent genes and the
probability of altruistic behavior is a non-monotone function of the number of
”altruistic alleles” in the genome (being the highest when this number
approaches some intermediate value). It was, however, recently suggested that
the probability of the origin and persistence of altruistic behavior was highly
increased in any sexual species where any behavioral trait, including the
altruistic behavior, is usually determined by a greater number of genes and many
of these genes have (due to epistasis) a context-dependent influence on the
particular trait. It was suggested that due to decreased heritability of traits,
the probability of persistence of altruistic traits in a population is highly
increased even for the altruistic behavioral patterns coded by several genes
with additive or semiadditive effects \cite{9}.

In the present study, we tested a verbal model \cite{9} based on the frozen
plasticity theory \cite{10} that suggests an increased probability of
persistence of the altruistic behavioral patterns when coded by several
substitutable genes rather than by a single gene with a large effect. The
present study starts with the description of the model.
 
\section{Model} 
We consider the \emph{fitness} in the classical meaning of the word, i.e. if two
individuals have the fitnesses equal to $a$ and $b$, then the ratio of the expected
number of their descendants is $a/b$. Especially, when the fitnesses of two
individuals are $1$ and $1+c$, we can say that the second individual has an
\emph{advantage} $c$ over the first one. It is easy to see that if the
fitnesses of two individuals are $a$ and $b$ and $a \leq b$, then the second
individual has an advantage $(b-a)/a$ over the first one.

Our model consists of a \emph{metapopulation} of $n \cdot m$ individuals.
They are structured into $n$ demes, of an average size of $m$.
We monitor the metapopulation's behavior in $N$ generations. Each generation
consists of three phases: natural selection, migration and mutation.

The phase of natural selection results in the replacement of all the individuals
by their descendants. Similarly as in $\cite{6,7}$ this happens ''at once'' and
the size of the metapopulation is preserved. Inside the metapopulation, two
kinds of natural selection take place, intrademic and interdemic. Under
intrademic selection, selfish individuals have an advantage $\alpha$ over
altruistic individuals, while under interdemic selection, altruistic demes have
an advantage $\beta$ over selfish demes. More formally, the meanings of
$\alpha$, $\beta$ are as follows: The probability of a new individual becoming
the member of the $i$-th deme is proportional to $m_i+\beta a_i$, where $m_i$ is
the size of the $i$-th deme and $a_i$ is the number of altruists of the $i$-th
deme.  The parents of this new individual are two randomly chosen individuals
from the $i$-th deme. The probability of a random individual becoming a parent
is proportional to its fitness which is equal to $1$ for an altruist individual
and to $1+\alpha$ for a selfish individual. The phase of natural selection is
ended by the extinction of any deme whose size is less than or equal to 2. Its
place is taken by the deme of the largest size, which randomly splits in two new
demes, with each individual being put randomly with a probability of $1/2$ into
the first or the second new deme.

Altruism is controlled by $gen$ genes of each individual. Each of these genes is
randomly inherited from individual's parents and each gene has two variants: an
altruistic and a selfish one. The phenotype of the individual 
depends on whether or not the number of altruistic alleles in the genome is
at least $thr$ (threshold).

Mutations occur randomly and for all generations, all individuals and any of
their genes the probability of an allele being switched from altruistic to
selfish or vice-versa is $\mu$. The individuals can migrate between demes and
the probability of an individual leaving its deme for another (randomly chosen)
deme (i.e. migration rate) is $\eta$. 

So far, the model has the following parameters: advantages of selfish
individuals and altruistic demes ($\alpha,~ \beta$), number of demes ($n$),
average deme size ($m$), mutation rate ($\mu$), migration rate ($\eta$),
altruism controlling mechanism ($gen$ and $thr$), number of generations used for
simulation ($N$) and the initial rate of altruists. By $p$ let us denote the
average rate of altruists during the whole evolution. We shall say that the
\emph{metapopulation is altruisic} if $p \geq \varepsilon$, where $\varepsilon$
is the minimal rate of altruists needed for a metapopulation to be considered as
altruistic. By $R$ let us denote such an advantage of the altruistic demes,
which leads to a $1/2$ probability of an altruistic metapopulation occurring.
More formally, $R$ is the advantage of the altruistic demes such that 

$$\P(p \geq \varepsilon)=\frac{1}{2}.$$

\noindent Naturally, $R$ should be seen as a function of all the parameters from
the model including $\varepsilon$. For practical usage, we shall always treat
$R$ as a function of just one parameter, other parameters will be fixed and
their values will be clear from the context. Under weak selection we can assume
that $R$, as a function of a selfish individual’s advantage $\alpha$, is
proportional to $\alpha$ ($R \sim \alpha$). Therefore, rather than $R$, we shall
analyze the fraction $r:=R/\alpha$ for low values of $\alpha$, in our case
$\alpha=0.01$. This value was chosen as a compromise between a too big $\alpha$
where $R \sim \alpha$ holds no more and a too small $\alpha$ where the effects
of both intrademic and interdemic selection are too subtle comparing to random
fluctuations of the system which enormously complicate the numerical analysis.

\section{Simulation}  
The model described above was straightforwardly simulated by a program written
in the C++ language. From $N=10000$ generations we calculated the average rate of
altruists $p$. If $p$ is sufficient (i.e. $p \geq \varepsilon$), we
decrease a little the value of $\beta$, and vice versa. By repeating this
procedure, we get $\beta$ oscillating around the searched value $R$. Sadly,
these oscillations and also the oscillations of averaged values of $\beta$ are
too big for calculating $R$ with sufficient accuracy. Therefore, we use another
approach: Instead of averaging the values of $\beta$, we attempt to find a linear
model that would ''explain'' the measured values of $\beta$ with some likelihood.
We search the space of all possible models for the \emph{best} model - such
model that best explains the measured values. This approach is well known as the
maximum likelihood method and is described in detail, for example,
in \cite{11}. The final value of $R$ is then deduced from the best model.

\section{Comparison of the model's behavior with the known theoretical results} 
 
The known theoretical results obtained by analytical computation generally refer
to a weak selection (even if it is not stated explicitly by some authors, their
assumptions only apply to weak selections). Therefore, they describe function
$r$ where $gen=1$ and $thr=1$. 

\begin{figure}[ht]
\centering
\includegraphics[width=0.6\textwidth]{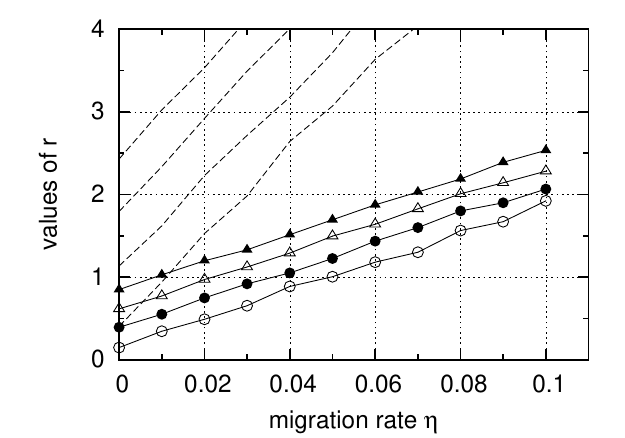} 
\caption{Values of $r$ as a dependence on migration rate $\eta$. Values of other
parameters: number of demes $n=10$, average deme size $m=10$,
$\varepsilon=1/2$. Mutation rate: $\mu=0.001$ (empty circles) $\mu=0.007$ (full
circles), $\mu=0.013$ (empty triangles) $\mu=0.019$ (full triangles). Dashed
lines show results for the same values of $\mu$ but for a different structure of the
metapopulation: $n=5$, $m=20$. The markings are approximately the size of the
$90\%$ confidence intervals calculated by the maximum likelihood method. 
}\label{fig:migracia} \end{figure}

Our first finding is that $r$ does not depend on the initial rate of altruists.
It clearly follows from the facts that there is enough time for alleles to
mutate in both ways, and since both $\alpha$ and $\beta$ are low, there is also
enough time for altruism to spread as a result of the genetic drift. Further
results are summarized in Figure \ref{fig:migracia}. In accordance with
\cite{1,2,5} we get linear dependency of $r$ on $\eta$ and in accordance with
\cite{1,5} we also get linear dependency of $r$ on $\mu$. In discordance with
\cite{1,5} the slope of the dependencies (i.e. $\frac{\partial R}{\partial
\eta}$) changes with different parameters $n$, $m$. The slope change is roughly
in agreement with \cite{2}, although there is an inaccuracy of about $30\%$ in
its actual quantity. Furthermore, the dependencies of $r$ on $n$ and $\mu$ are
neglected in \cite{2}. 

The disagreements mentioned above are a consequence of too strong assumptions
taken by some authors in deriving their results. As an example, the
stochasticity of the system is neglected \cite{2}, only demes
containing either only altruistic or only selfish individuals are reproduced \cite{1} or there
is a neglect of strong correlations between dependent random variables \cite{8}. We
do not want to belittle any of the papers cited. Quite the opposite, our point is
to illustrate that solving this problem analytically is very hard and resists
numerous different approaches, no matter how inventive they are in using a wide
range of mathematical tools. 

\section{Weak selection and multiple gene control} 

By the weak selection we mean a regime where the fitness of all individuals is
close to one. On the other hand, the strong selection means that the fitness can
significantly vary between individuals. Naturally, these terms are not strictly
defined and there is a continuous scale between the weak and strong selection. 

The results from the simulation for some combinations of the parameters are presented in
Tables $\ref{tab1}$ and \ref{tab2}. All $r$ values in Table $\ref{tab1}$ are very
close to each other. It is not accidental as we explain below. 

Let us assume a metapopulation with the same rate $p$ of the altruistic alleles in
each deme and on each gene. The rate of the altruists ($f$) is then
function of $thr, gen$ and $p$. Especially for $thr=1$, it can be calculated
as follows:
\begin{equation}
\label{frek}
f:=1-(1-p)^{gen}
\end{equation}

\noindent (this is the case for models 1,2,3,5,6 and 7). For odd $gen$ and
$thr=\ceil{gen/2}$, the frequency of the altruists $f$ equals $p$ (the case for
models 4,8). We can say that allele of gene $g$ is \emph{active} if its change
(from altruistic to selfish or vice versa) changes the phenotype of the
individual. It clearly happens if (and only if) there are exactly $thr-1$
altruistic alleles among the rest of the genes. Probability of such event will
be denoted as $P=P(p,gen,thr)$. Let us estimate the advantage that is conferred
by a selfish/altruistic allele to the individual/deme, respectively. With a
probability $1-P$, this allele does not influence anything and therefore, does
not confer any advantage to anyone. With a probability $P$, the allele confers
advantage $\alpha$ to the individual (if the allele is selfish) or advantage
$\beta$ to the deme (if the allele is altruistic). We can regard the average
advantages of the allele as the product of the probability of the allele being
active multiplied by the advantage conferred to the individual/deme. We conclude
that the selfish allele confers average advantage of $P \cdot \alpha$ to the
individual and altruistic allele confers advantage $P \cdot \beta$ to the deme
it inhabits. As the assumption about the equal distribution of the altruistic
alleles holds, we can analyze the evolution of the gene $g$ alleles
independently on the evolution of other alleles of other genes. Let us now
compare models 1 and 3. In model 1, we found a ratio $r_1=\beta / \alpha$ which
leads to a rate of altruists (and therefore as well to a rate of altruistic
alleles) $50\%$. If the same values $\alpha, \beta$ are used in model 3, then
the ratio of the actual advantages the allele confers in this model is: 

$$r_3=\frac{P \cdot \beta}{P \cdot \alpha}=\frac{\beta}{\alpha}=r_1.$$ 

\noindent Since we are in the regime of low values $\alpha$ and $\beta$, we only
expect the ratio of $\alpha$ to $\beta$ (and not the absolute values of $\alpha$
and $\beta$) to be important for the rate of the altruistic alleles. This ratio
does not change ($r_1=r_3$) and therefore as in model 1, we also expect $50\%$
of altruistic alleles on average also in model 3. By using equation \ref{frek},
we get the final rate of altruists $31/32$, which is exactly the value of
$\varepsilon$ in model 3. Using a similar approach, we can justify the
similarity of all other results in Table \ref{tab1}. In Table \ref{tab2}, we
work with lower values of $\eta$. This causes that the altruistic alleles cease
to be evenly distributed and start grouping in some demes that will become more
altruistic. It means that the results derived in Table \ref{tab1} also cease to
hold. 

Summary: Under weak selection and strong migration (10\% in our models),
multiple gene control does not significantly affect the $\beta/\alpha$ ratio
needed for altruistic alleles to occur. On the other hand, multiple gene control
does significantly affect the final rate of altruists (Table \ref{tab1}).

Under weak selection and not so strong migration, multiple gene control also
helps to spread altruism by changing the $\beta/\alpha$ ratio (Table
\ref{tab2}). Although we are not able to explain such an outcome sufficiently,
the next paragraph provides at least a partial explanation of this phenomenon.

\begin{table}
\caption{Models' parameters: $n=20, m=30, \mu=0.001, \eta=0.1$}
\label{tab1}
\begin{center}
  \begin{tabular}{|| c | l | l @{ = } r || }
    \hline \hline
    Model 1 & $gen=1,~ thr=1,~ \varepsilon=1/2$ & $r$ & $5.2 \pm 0.1$ 
    \\ \hline
    Model 2 & $gen=3,~ thr=1,~ \varepsilon=7/8$ & $r$ & $ 5.6 \pm 0.1$ 
    \\ \hline
    Model 3 & $gen=5,~ thr=1,~ \varepsilon=31/32$ & $r$ & $5.7 \pm 0.3$ 
    \\ \hline 
    Model 4 & $gen=5,~ thr=3,~ \varepsilon=1/2$ & $r$ & $5.7 \pm 0.1$ 
    \\ \hline \hline
  \end{tabular}
\end{center}
\caption{Models' parameters: $n=20, m=20, \mu=0.0005, \eta=0.01$}
\label{tab2}
\begin{center}
  \begin{tabular}{|| c | l | l @{ = } r || }
    \hline \hline
    Model 5 & $gen=1,~ thr=1,~ \varepsilon=1/2$ & $r$ & $0.50 \pm 0.03$ 
    \\ \hline
    Model 6 & $gen=3,~ thr=1, ~ \varepsilon=7/8$ & $r$ & $ 0.43 \pm 0.03$ 
    \\ \hline
    Model 7 & $gen=5,~ thr=1,~ \varepsilon=31/32$ & $r$ & $0.24 \pm 0.03$ 
    \\ \hline 
    Model 8 & $gen=5,~ thr=3,~ \varepsilon=1/2$ & $r$ & $0.47 \pm 0.03$ 
    \\ \hline \hline
  \end{tabular}
\end{center}
\end{table}

Let us compare Model 5 and Model 7 from Table \ref{tab2}. Let us start with
Model 5 and assume a purely altruistic metapopulation where one selfish migrant
emerges in one of the demes. It spreads in the deme which consequently shrinks
and finally vanishes. Until the extinction it spreads selfish migrants and
potentially infect other demes by selfishness. It may happen that the selfish
migrant or its descendants will not reproduce although they have advantage
$\alpha$ over the rest of the deme. It would change if altruism was controlled
by five genes as it is in Model 7. Since $thr = 1$, even the descendants of a
selfish migrant are with high probability altruists. Although the selfish
migrant brings some selfish alleles to the new deme, their spreading is
influenced mostly by random drift and they do not gain significant advantage
from the selfish behavior. 

\section{Strong selection and multiple gene control} 

In \cite{3} authors showed that under sufficiently low migration and
mutation rates, the metapopulation can exist in two different semi-stable states.
They are called S state, with almost all individuals being selfish, and A state,
with almost all individuals being altruistic. Transitions between these two
states are denoted as A-S and S-A. 

The results of our simulation are in full accordance with existence of S and A
states described in \cite{3}. Although the authors of \cite{3} used extremely low
values of $\mu$ and $\eta$, we observed the S and A states also when higher
values of $\mu$ and $\eta$ together with higher advantages $\alpha$ and $\beta$
were set. Note that naturally there is a strong correlation between existence of
S and A states and the fact that $R$ depends on the initial frequency of
altruists. 

In our simulations, we focus on A-S transition, i.e. we initialize all alleles
in the metapopulation as altruistic. A similar shape of $R$ as a function of
$\beta$ is observed for S-A transitions, but higher values of $\beta$ are
necessary for S-A transitions. 

\begin{figure}[ht] \centering
\includegraphics[width=0.49\textwidth]{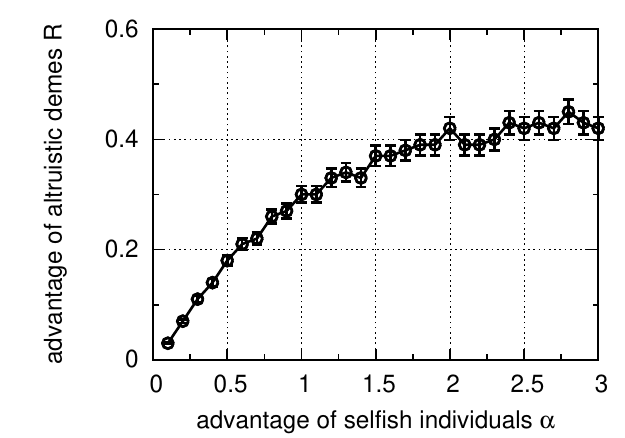} \hfill
\includegraphics[width=0.49\textwidth]{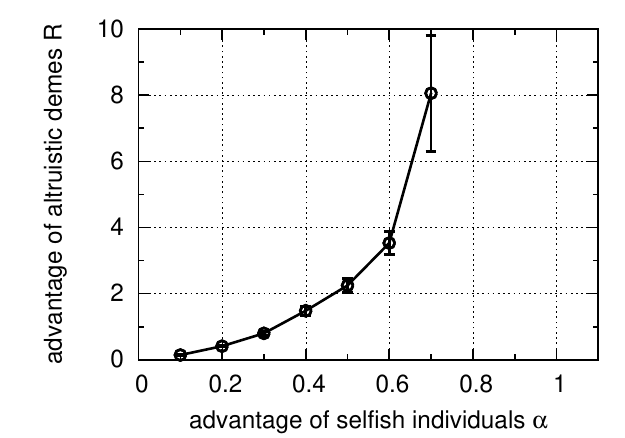} \caption{Values of $R$
as a function of $\alpha$. The horizontal asymptote can be observed on the left,
the vertical asymptote on the right. Parameters of the model: The number of demes
$n=10$, average deme size $m=10$ (left) or $m=30$ (right), mutation rate
$\mu=0.001$, migration rate $\eta=0.01$.} \label{nabeh} \end{figure}

We will start with a brief explanation of the behavior of $R$ as a function of
$\alpha$ under one-gene control of altruism. For small $\alpha$ function $R$
is similar to linear dependence, but for bigger $\alpha$, this no longer
holds, and the function gets concavely or convexly shaped. Finally the $R$ tends
to asymptote either in horizontal or vertical direction. The horizontal
asymptote is shown in Figure \ref{nabeh} on the left. It means that a relatively
small advantage for altruists is sufficient for compensating for a much bigger
advantage of selfish individuals. The vertical asymptote is shown in Figure
\ref{nabeh} on the right. The interpretation is that a particular advantage of
selfish individuals cannot be outweighed by any advantage of altruists, no
matter how high. 

We briefly describe the mechanism how different asymptotes occur. Let us suppose
$\alpha \rightarrow \infty$, a relatively small $\beta$ and a purely altruistic
metapopulation where one selfish mutant emerges. This mutant spreads over its
deme very quickly. The deme starts shrinking and finally goes extinct. If it
succeeds in producing enough selfish migrants that spread, the selfishness will
thrive. A similar effect occurs when $\beta \rightarrow \infty$ and $\alpha$ is
relatively small. This time the selfish mutant spreads slowly in its own deme
that is also shrinking proportionately slowly because of a lower number of
altruists in it. Once there are no altruists in the deme, it almost immediately
goes extinct. Once again, the main factor that influences the spread of the
selfishness is the number of selfish migrants that are produced by the infected
deme until it goes extinct. The number of selfish individuals is significantly
influenced by the quantity $m \cdot \beta$ which says how many migrants are
produced in one generation by an average sized deme. It is also influenced by
values $\beta$ (for horizontal asymptotes) and $\alpha$ (for vertical
asymptotes), since these values predict how fast the deme will be shrinking.
Contrary to this, no matter how big the values $\alpha$ (for horizontal
asymptotes) or $\beta$ (for vertical asymptotes) are, a selfish individual
cannot spread over its deme faster than in one generation and also a deme
consisting of only selfish individuals needs at least one
generation to extinct. Therefore, it does not matter how high these quantities
are, once they are high enough, their change does not influence the number of
selfish migrants. The asymptotic behavior is therefore a natural consequence of
the systems being insensitive to a change of some parameters. 

\begin{figure}[ht] \centering
\includegraphics[width=0.49\textwidth]{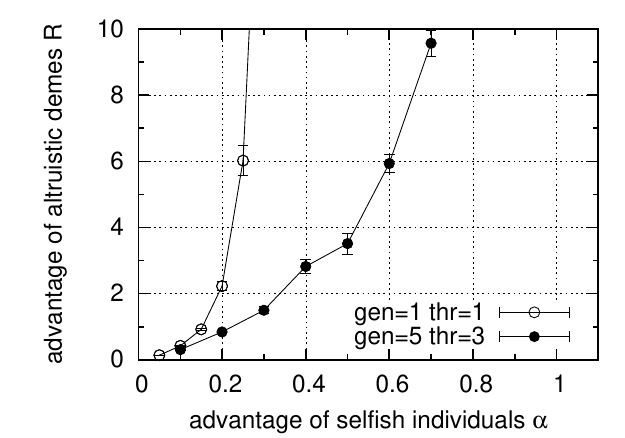} \hfill
\includegraphics[width=0.49\textwidth]{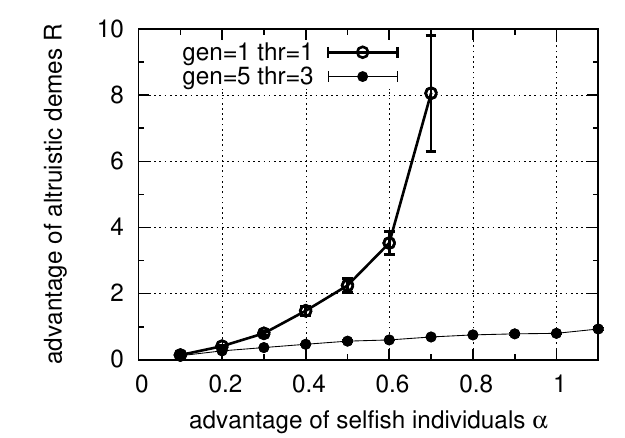} \caption{Values of $R$
as a function of $\alpha$. Left: average deme size $m=55$. Vertical asymptote
also occurs for $gen=1, thr=1$ and for $gen=5, thr=3$. Right: average deme size
$m=30$, vertical asymptote for $gen=1, thr=1$ (empty circles), but horizontal
asymptote for $gen=5, thr=3$ (full circles). The other parameters are equal to
those given in Figure \ref{nabeh}: $n=20$, $\eta=0.01$, $\mu=0.001$}
\label{konvex_nabeh_55} \end{figure}

Let us now introduce the results for multiple gene control. We put $gen=5$ and
$thr=3$. Let us remind that under the weak selection, there was no difference
between $gen=5$, $thr=3$ and $gen=1$, $thr=1$. Differences now occur as an
effect of strong selection and existence of semi-stable S and A states. Values
of $R$, as a function of $\alpha$, are presented in Figure
\ref{konvex_nabeh_55}. The figure on the left demonstrates that under five-gene
control, the observed values of $R$ are significantly lower than under one-gene
control. Also, the placement of the vertical asymptote is different. For
one-gene control, its $x$ coordinate is almost three times as high as that for
five-gene control. It means that under five-gene control altruism can occur,
even for such $\alpha$ that makes it completely impossible under one-gene control. If
we reduce the average deme size to $m=30$, an interesting phenomenon occurs: for
one-gene control, we still have vertical asymptote, while for five-gene control,
we obtain horizontal asymptote. It is possible due to the fact that $m \cdot
\eta = 0.3$ which is roughly close to one. 

We conclude, that under five-gene control, metapopulation can be altruistic even
for such low values of $\beta$ that would lead to complete annihilation of
altruism under one-gene control.

\section{Discussion} 
Although quite unfavorable conditions were chosen for altruism to spread (unlike
others we used higher mutation and migration rates), the advantages of
altruistic demes needed for altruism to spread are rather low. Even for one-gene
control, these values are lower than predicted by other theoretical studies.
This is specific for the discussed model where altruistic demes thrive and
produce other altruistic demes while selfish demes shrink and therefore produce
fewer migrants. When intrademic selfishness only results in
higher probability of deme extinction (but until extinction the size of the deme
is constant, as is the case for example in \cite{1}, which presents a model that
appears to be less realistic than ours), the values of $R$ are significantly higher. 

The model for multiple gene control of altruism used in this work is absolutely
symmetric - an altruistic allele has exactly the same phenotype effects as any
other altruistic allele present in the genome. In the reality, the situation
will be different, with the phenotype effects of two altruistic alleles on two
different genes varying from one another. Preliminary results of simulations of
this scenario indicate that even this way of controlling altruism is quite
efficient for spreading it. 

Another important property of this model is that having one more altruistic
allele always makes the individual at least as altruistic as it was before. If
some negative dependencies were considered (i.e. having one more ''altruistic''
allele could result in a more selfish individual), then the spreading of
altruism would be much easier. The important message of this paper is that, in
agreement with the prediction of the former verbal model \cite{9} based on the frozen
plasticity theory \cite{10}, even quite trivial dependencies among altruistic alleles
may have a strong impact on the spreading of altruism. 

\vskip0.5cm
\noindent \textbf{Acknowledgements.} 
The first author was supported by VEGA grant No. 1/0406/09. The second author
was supported by grant No. 0021620828 from the Ministry of Education, Youth and
Sports of the Czech Republic.

\end{document}